# Influence of crystalline structure on RF dissipation in Niobium: flux trapping, hydride precipitate, doping behavior...

C. Z. Antoine

*CEA, Irfu, SACM, Centre d'Etudes de Saclay, 91191 Gif-sur-Yvette Cedex, France*

E-mail: Claire.antoine@cea.fr

## Contents

**Abstract.**

Bulk niobium is the material mostly used in RF superconducting cavities for accelerator. Predicting and reducing the surface dissipation in RF is mandatory, since it has a tremendous cost impact on most of the large accelerator projects. The theoretical approach of superconducting radiofrequency (SRF) behavior has been far less explored than DC behavior and is still based on the description of relatively simple systems, quite remote from the realistic material in use. Nevertheless, because the actual crystalline substructure is not taken into account, it is still difficult to predict surface dissipation accurately. Moreover, Niobium with its large $\xi$ (~40 nm), exhibit an original behavior compared to the usual superconductors used in applied superconductivity, and generalities (e.g. pinning at grain boundaries) needs to be reconsidered.

In this paper we hope to demonstrate that sources of dissipation usually attributed to external causes (mainly flux trapping during cooldown and hydrides precipitates) are related to the same type of crystalline defects which are not grain boundaries; that they affect the local superconducting properties and can also be at the source of early vortex penetration at the surface. We want also to stretch out how those defects can explain some of the discrepancies observed from lab to lab on recent results obtained i.e. in doping experiments. Understanding the origin and the role of these defects could provide indications for updating specification as well as fabrication follow up.

Rather than new results the author wishes to present to the superconducting accelerators community the synthesis of multiple experimental results scattered in the literature that provide a new lighting on recently published work, in particular in the domain of SRF cavity doping as well as issues such as sensitivity to trapped flux during cooldown.

We will also try do draw whenever possible some conclusion about other types of superconductors used for SRF applications, including Nb/Cu thin films and to discuss the possible change of behavior with field or frequency. We will concentrate on surface and material science aspects since the experimental results on RF cavities have been already treated elsewhere (see e.g. [1]).

## 1. Introduction:

Most of the present models that try to determine the surface resistance of superconductors in the RF regime (SRF) do not take into account most common crystalline defects found in bulk niobium. Primary models that allow to evaluate

RF surface resistance (Mathis and Bardeen integrals) are based on the 2-fluids model that supposes that the density of superconducting electron $n_s$ is uniform. As has been amply demonstrated during the development of superconducting (SC) material for magnet, a lot of phenomena are related to the fact that $n_s$ is not uniform [2]. Ginsburg Landau developments which account for some non-uniformity and non-linear behavior are abundant in the literature but they are valid only close to $T_C$ or $H_{C2}$, unless one deals with linearized GL model which applies at any T, but only for high ξ. Indeed the majority of the numerical developments found in the literature have been conducted for high κ superconductors and their numerical predictions are generally poor for Nb in SRF operating condition (Meissner state, 2-4K) [3, 4]. Recent models proposed inside the RF community deal with defects and some non-uniformity, but most of the time defects are considered to be randomly distributed, which is not a fully realistic description [5-7].

### 1.1. Crystalline defects

From the metallurgical point of view, defects like dislocation or interstitial atoms do not distribute evenly inside the lattice, but tend to appear and/or diffuse along specific crystalline orientations [8]. Moreover since some defects interact together, they tend to gather in specific crystalline regions, e.g. at grain boundaries, or at sub-grain levels. In the following, we hope to demonstrate that the same type of defects is at the origin of several phenomena of interest, that are usually considered as external to the bulk material, such as field trapping or the apparition of hydride precipitates during cooldown. Most of these defects induce an elastic deformation of the lattice (Figure 1), which can be quantified in some case, and that has a large impact on the atomic diffusion behavior such as interstitial atoms diffusion implied in doping, or dislocation movement implied in deformation and/or annealing effects. Those defects are as well at the source of the local variations of the SC parameters, including the local density of Cooper pairs $n_s$ and thus on their pinning behavior (see below for more details).

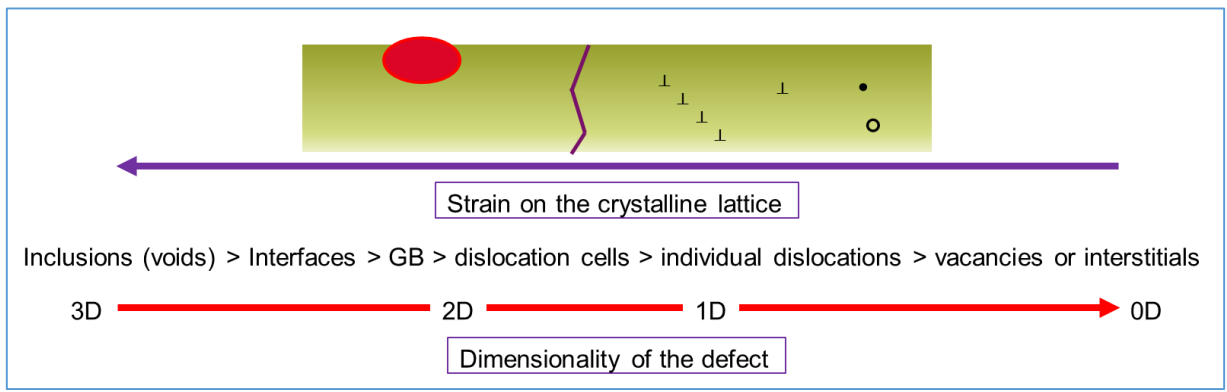

*Figure 1: Typical defects found in crystalline structures. The strain increases with dimensionality of defects; it affects atomic diffusion behavior and also affect local SC properties like ξ or mean free path as well (see text below for more details).*

### 1.2. Generalities on pinning;

Pinning behavior of crystalline defects has been thoroughly studied from the theoretical point of view during the development of SC cable (see texbooks, e.g. [3, 4, 9]); a lot of data can be found on Nb, bulk as well as thin films in the literature (e.g. [10-14]). As proposed before, the detail of the physical phenomena implied in the process can also be used to explain other type of behavior observed on cavities.

Matsushita [4] describes four main mechanisms involved in pinning behavior. The relative weight of each mechanism depends on the defect types as well on the SC properties of the considered material.

A. Condensation energy gain. Making a vortices go through a normal conducting area (or at least an area with depressed superconductivity) results in a gain in Gibbs free energy. At equilibrium it is counterbalanced by the energy necessary to elastically deform the vortices lattice.
B. Elastic interactions. The moduli in the superconducting state are always less than in the normal state [15]. Moreover ΔY/Y, the relative elastic difference between normal and SC state, is known to decrease with mean free path decreasing [11]. Since crystalline defects also modify the local elastic properties of the lattice, one can expect strong interaction between the two mechanisms in some cases.
C. Magnetic interaction. If defect's size is larger than λ, you can treat it like an interface with image vortex formalism and the appearance of a surface barrier.
D. Kinetic interaction. Areas with different ξ exhibit different vortex velocities, resulting in local effects that are not taken into account in the simple London model (2 fluid approach).

In a general way, pinning results from local effects and local variation in the SC parameters, and there are several ways to achieve efficient pinning:

- "Surface magnetic pinning" where the main mechanism is magnetic pinning resulting from large defects: twinning, voids, non SC aggregates, irradiation defects (columnar), nano-indentations. In that case dominant mechanism are A (Condensation energy gain) and C (Magnetic interaction). As can be intuitively guessed, the maximum pinning force occurs when the dimension of the defect is close to $2\xi$.
- "Volume core pinning" where one weak pinning centers, but many of them (e.g. interstitials or small precipitates evenly distributed in the material, isolated dislocations). The main mechanism would be B (Elastic interaction), which much less efficient than the magnetic interaction, but if they are many pinning centers it results into very strong pinning nonetheless.

## 2. Pinning in Niobium

### 2.1. General observations

Pinning is at the origin of irreversibility in DC magnetization curves, as the surface inside the hysteresis loop of a magnetization curve is directly proportional to pinning force [16]. As can be seen on Figure 2, irreversibility is also related to crystalline state of the material, and recrystallization improve greatly the reversibility of the magnetization curve. Effects of damage layer and/or alloying on AC losses of Nb have been explored in [17] where it is shown that AC losses can be directly inferred from the DC magnetization curve. Note that in some case pristine surface exhibit a lot of flux jumps characteristic of surface pinning due to damage layer [18]. Nevertheless, even after a 1400°C treatment some pinning still occurs, mainly due to pinning on the surface and edges of the sample [19]. Since high temperature treatments have a detrimental effect on mechanical resistance of bulk Niobium, one has to consider that some flux trapping will always occur and chose a compromise for the heat treatment. Therefore it is also important to determine which kind of crystalline defects is more detrimental to SRF applications.

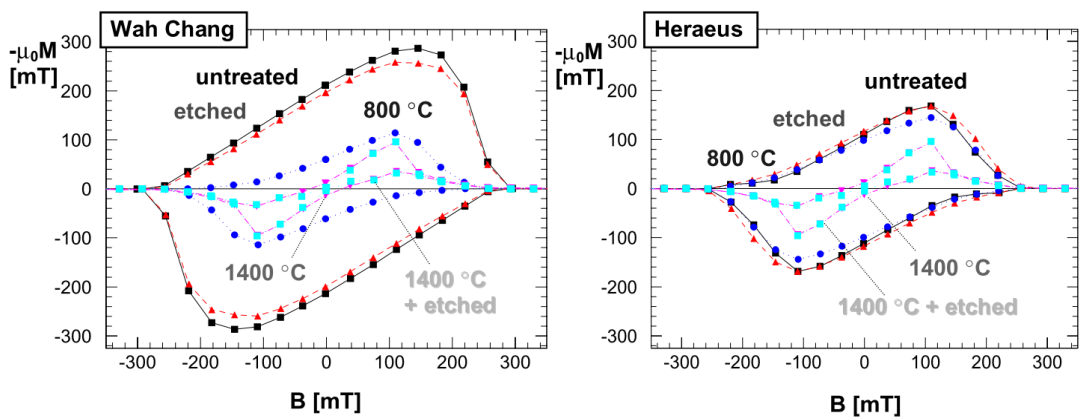

Figure 2: DC magnetization curves from two different Nb batches (from [20]). One can observe that recrystallization at 800°C and even better at 1400°C decreases the hysteresis in the curve, indicating a decrease in pinning efficiency. Nevertheless it is not possible to make totally disappear, especially within this type of measurement configuration where surface pinning and edge pinning always occur.

### 2.2. Surface pinning

Surface pinning was not considered until recently in SRF, but they are several experimental indications that it might play an important role. For instance Pautrat et al studied the surface pinning on surface corrugated Nb samples in relation to roughness [21]. Roughness can be characterized by a distribution of angles determined by Fourier transform of the autocorrelation function of the surface roughness measurement. He showed that increasing roughness increases the critical current (which is proportional to the amount of flux pinning), but he also showed that since the distance between vortices change with field, the range of roughness to be considered changes too. Indeed, at low field one is mainly sensitive to defects around 0.1-1 μm while at higher field one is sensitive to smaller scale roughness. It is believed that roughness far below the $\xi$ range does not play a role. The explanation for such a high effect of surfaces/interfaces results from the limit condition: $\vec{J_s} \times \vec{n} = \vec{0}$, which obliges the vortices to exit the superconductor always perpendicular to the surface. If the surface present a certain roughness, it results into bending of the vortex, which increases its length, which in turn increases the elastic return force. As pinning and elastic forces are in equilibrium, hence the pinning force also automatically increases. This effect is mainly observed on thin films where the surface/volume ratio is high, and it could play a paramount role for alternative superconductors (NbN, $Nb_3Sn$, $MgB_2$...) in SRF application: indeed most of these material are fabricated into relatively thin films with relatively small grain sizes at the origin of some roughness at the nm level. One has still to verify that the impact of surface nano-roughness keeps moderate in SRF applications.

A similar rationale applies to edge effects in magnetometry [21-23]. It might also play a role for nearly perfect material where little or no volume pinning occurs and probably can explain some of the µSR results observed recently on non-masked samples (e.g. in [24] and below).

*2.3. Effect of frequency: bulk vs thin films*

A very simple technique is widely used to determine volume pinning force [25, 26]; it consists in measuring complex penetration depth $\lambda_{AC} = \lambda' + i\lambda''$ in a AC set-up as shown in Figure 3. The measured flux is proportional to $\lambda_{AC}$: $\phi_{ac} = \int b_{ac}\, dS \sim 2\lambda_{ac}\ell_b b_0$.

At low frequency, $\lambda''$ is proportional to $\delta_{RF}$ (penetration depth in normal state) since the flux fully enters the superconductor and $\lambda' \sim 0$ (magnetic screening is inefficient). Since the vortices are efficiently pinned at low frequency, the resistance Rs is strictly 0 (so-called Campbell regime).

At high frequency pinning is inefficient and $\lambda' \sim \lambda'' \sim \lambda_L \ll \delta_{RF}$. The surface resistance follows the usual $R_{BCS} + R_{Res}$ behavior. The maximum of $\lambda'$ is proportional to pinning force and defines the so-called depinning frequency. Note that critical currents are also a good way to evaluate effective pinning force.

Figure 4 shows the results for bulk monocrystalline high quality Nb after [25]. The depinning frequency lies around $10^4$ Hz which shows that it is very unlike that pinning still persists at SRF frequency.

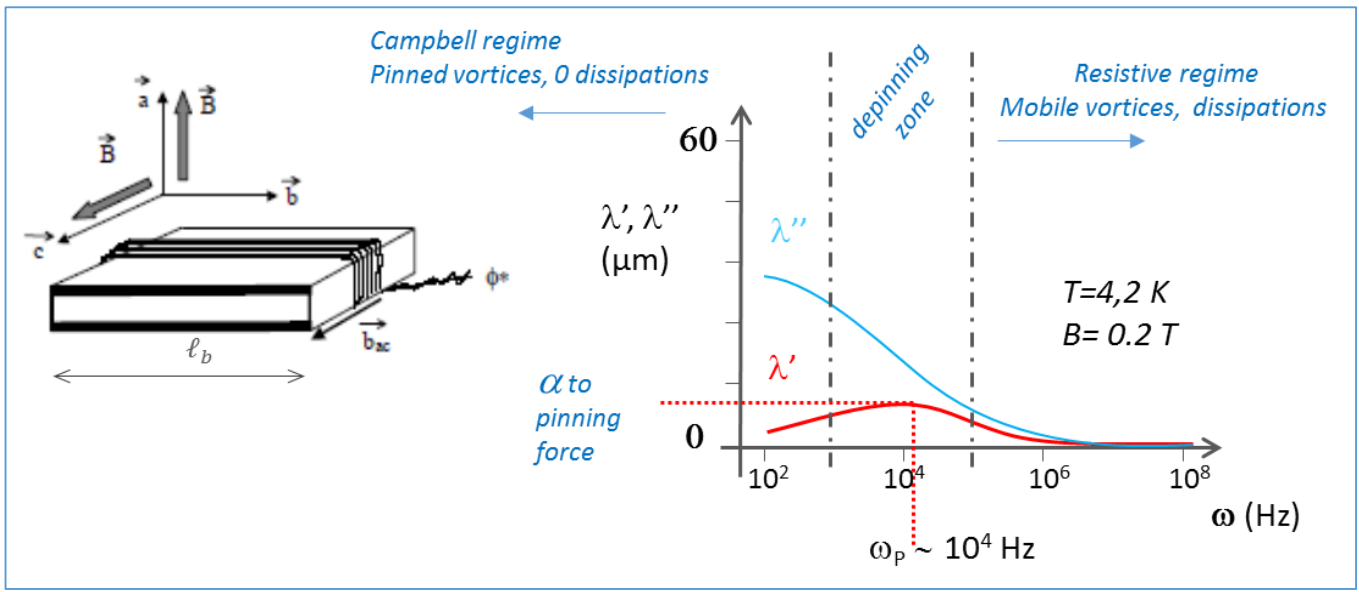

*Figure 3: principle of measurement of Ac penetration depth and results obtained for high quality Nb monocrystal (after[25])*

The same type of experiment applied to thin film Nb deposited with various techniques show that depinning frequency varies a lot with the quality of the films, but is typically ranging in the 10-30 GHz interval [27-30]. This explains why thin films show low sensitivity to trapped magnetic flux: all the flux is efficiently trapped and those superconductors keep in the Campbell regime. See also [31, 32] for theoretical support and [33] for recent SRF results.

In the next paragraph we will describe generalities about flux trapping and in the following paragraph we will describe crystallographic defects commonly found inside Nb and try to determine the main relevant pinning mechanism for SRF grade Niobium.

*2.4. Flux trapping during cooldown*

Flux trapping originates mainly from the trapping of residual (earth or thermo-current induced) magnetic field when cooling down of the cavities. Its study started in the 90's, when it was showed that thin film cavities and bulk behaved differently, thin film being apparently insensitive to external magnetic flux [34]. During his PhD (1994), C Valet showed for the first time that cooling down in field ranging from a few mG to some hundreds mG, 100% of the field was trapped [35-37]. He explored various aspect (RRR, effect of frequency, thermocurrents, presence of magnetically active component etc…). His work resulted in a better magnetic shielding of experimental cryostats, and a special care to prevent use of magnetized component. Altogether it allowed to improve mean Q values by a factor 10 allowing for the first time to get values above $5.10^{10}$ [38]. After a period of improvement of magnetic shielding and cryostat design, field trapping was somewhat forgotten since most of the projects in the 90's-00's were more concerned by high gradients. For instance in a project like TTC, the temperature regulation was such that most of the time the static losses of the cryostat where dominated by the BCS component, as the temperatures reached sometime 2 to 2.2 K at the end of the cryostats [39]. The conception of the magnetic shielding was sufficient to reduce the residual part in those conditions [40]. Nevertheless some improved model have been proposed like e.g.[32, 41, 42]. In [41], Ciovati showed that it was possible to move the hot spots and/or decrease their induced losses by the mean of a local heating (lamp), a scenario well-fitting the depinning of flux line with sufficient activation energy. Nevertheless it was not effective on all hot spots.

Recently the topic became hot again, because in many recent projects, especially for CW machine, decreasing the cryogenic losses by a factor 2 or 3 leads to significant savings from the point of view of infrastructures as well as operating costs. Many thorough study have been published recently (see e.g. [33, 43, 44]), showing the importance cooling rate, thermal gradient, influence of thermocurrents on dressed cavities, but no complete consensus appears: in some case high thermal gradients help to expel the field while in some other a small gradient is preferable, differences have been observed between vertical and horizontal cavities, dressed vs bare cavities. It seems that the results are still deeply impacted by some local variations, either in the local disposition of the labs or in the quality of the Niobium and/or of the cavities that have not been yet understood.

Flux trapping and imperfect expulsion of the flux at decreasing field is closely related to pinning. As the details of SRF experiments have been presented elsewhere [1]; in the following we will concentrate of material aspects of pinning, in particular on the occurrence and behavior of specific defects.

## 3. Typical features of the Niobium crystal structure:

### 3.1. Grain boundaries

Grain boundaries (GB) and interfaces are 2D defects where the lattice accommodates the differences in atom spacing between two adjoining grains with different orientation (or even different crystal lattice in the case of the oxide –metal interface). It results into lattice distortion that strongly depends on the relative orientations of the grains. GB are usually expected to be effective vortex traps [4, 10]. Indeed the pinning on GB is most effective when the crystalline disorder around the grain boundary extends in the same order of magnitude as $\xi$. This is the case in thin film superconductors, where disorder and grain boundaries density are high. Grain boundaries play also a strong pinning role like e.g. high Tc SC. Compounds superconductive only in a narrow range of stoichiometry (e.g. $Nb_3Sn$) might also be in this case, especially since they have a small $\xi$.

SRF quality Nb exhibit equiaxial structure and grains diameters ranging from 50 μm to several mm (or even cm for large grains); the distortion around GB does not extend more than a few atomic layers whereas $\xi$ is ~40 nm, i.e. much larger than the distorted region at GB. So grain boundaries in SRF grade Nb are not expected to provide strong pinning centers.

Several experiments demonstrated indeed some weakening of superconductivity at GB. In particular early field penetration can be observed in GB when exposed to magnetic field, forming Josephson vortices with penetration field ~1/3 of the bulk value [10, 45-48], the effect being maximum when the field is parallel to the field plane [3, 45]. Dasgupta [10] worked on bicrystal and evaluated the pinning force to be ~$7.10^{-6}$ N/m for a GB. One also observe that large grains material tend to exhibit a slightly higher mean field Q level than small grain material [18, 49]. Nevertheless, although grain boundaries are obviously areas of weakened superconductivity in bulk Nb, they do not affect particularly SRF behavior: hot spots nor quenches are seldom found at grain boundaries, and monocrystalline cavities exhibit qualitatively the same behavior as polycrystalline cavities [50].

### 3.2. Dislocations

Dislocation are linear defects that are always present in non-perfect crystals (equilibrium concentration depending on temperature). Upon deformation more dislocations can be emitted to accommodate the strain. In heavily deformed metals dislocation density can reach $10^{11}$-$10^{12}$ cm$^2$, but even a very carefully recrystallized material still hold dislocations even at temperatures close to the melting point. Isolated dislocations are expected to have a smaller pinning energy than 2D defects: the width of a dislocation is small, displacements being comparable with the interatomic distance and confined to a few atoms around the dislocation. Typically, for Nb the pinning force per unit length $F_p$ is about $2.10^{-6}$ N/m for an isolated dislocation (to be compared to $7.10^{-6}$ N/m for a GB [10] and $9.1\ 10^{-4}$ N/m for a normal conducting precipitate [4]). But they are many more dislocations than grain boundaries in typical Nb used for SRF applications.

Moreover, most of the time dislocations come into dislocation cells where dislocations pill up into "walls" with high stress concentration and high stress gradients whereas the inner cell are nearly defect less. For heavily deformed material they prefigure future grain or sub-grain boundaries that will appear upon recrystallization. Dislocation structure becomes non uniform in Nb for very small amounts of deformation (3-5 %, equivalent to a density of dislocation ~$10^8$-$10^9$ cm$^2$) [3]. Examples of the effect of recrystallization on density of dislocation in Nb can be found e.g. in [49, 51]. Those dislocation cells are clearly 2D defects similar to GB, rather than linear, and their spatial extension is much larger than the distortion around GB. So they clearly are a better candidate for pinning center in Nb than GB. This has several other consequences.

Dislocations do not appear randomly; they move along slip planes and within the slip planes, there are preferred crystallographic directions for dislocation movement (slip directions), typical of a BBC lattice for Nb [8, 52]. Which means that during forming a polycrystalline material, not all grains deform the same way depending on their initial orientation (this phenomenon is at the origin of the apparition of orange peel when forming metals). Because of an "unfavorable" orientation some grains will see their density of dislocation increase a lot, and they might end with a specific texture which is hard to recover even after a prolonged recrystallization. This aspect is particularly strong after cold rolling where the surface texture has shown to be resistant to recrystallization [53]. Materials with severe near surface deformation exhibit a thin surface layer able to support current densities up to 2 orders of magnitude higher than the bulk values [3], which is a good indication of the existence of surface pinning. This scenario explains very well the patchy nature of hot spots (see §5), but also why even after recrystallization some grains retain a high density of dislocations. For instance Hering [54] found a density of dislocation as low as $10^6$-$10^7$ cm$^2$ in monocrystalline, RRR 3000, well recrystallized Niobium. He evidenced pinning with a decoration technique showing the deformation of the Abrikosov lattice combined with TEM inspection. In the recrystallized material very few pinning occur, whereas the same material after cold work exhibits strong pinning, with vortex lattice destroyed at large scale. Short scale ordering can still be observed inside the dislocation cells, but stops at the cell walls All vortices lines at short distance from the wall get attracted there. Similar effects have been observed in [12, 51]. In [12], Sathanam drew a correlation between $J_C$ (which is proportional to the pinning force $F_P$) and the mean size of dislocation cell. He found out that $F_P$ is maximum for dislocation cells about 100 nm, which is precisely the same order of magnitude as 2ξ. It is quite interesting that Martinello found that for cavities maximum trapped flux sensibility was also centered on the same 100 nm value for mean free path [55, 56]. On the other hand, similar experiment made at Cornell showed a maximum trapped flux sensibility for niobium with a *m.f.p.* centered around 8 nm [42], showing that they are still non reproducible elements involved in the fabrication of the cavities from lab to lab. Such a short mean free path seem to indicate that the Nb of the Cornell cavities is rather dominated by point defects (see next §).

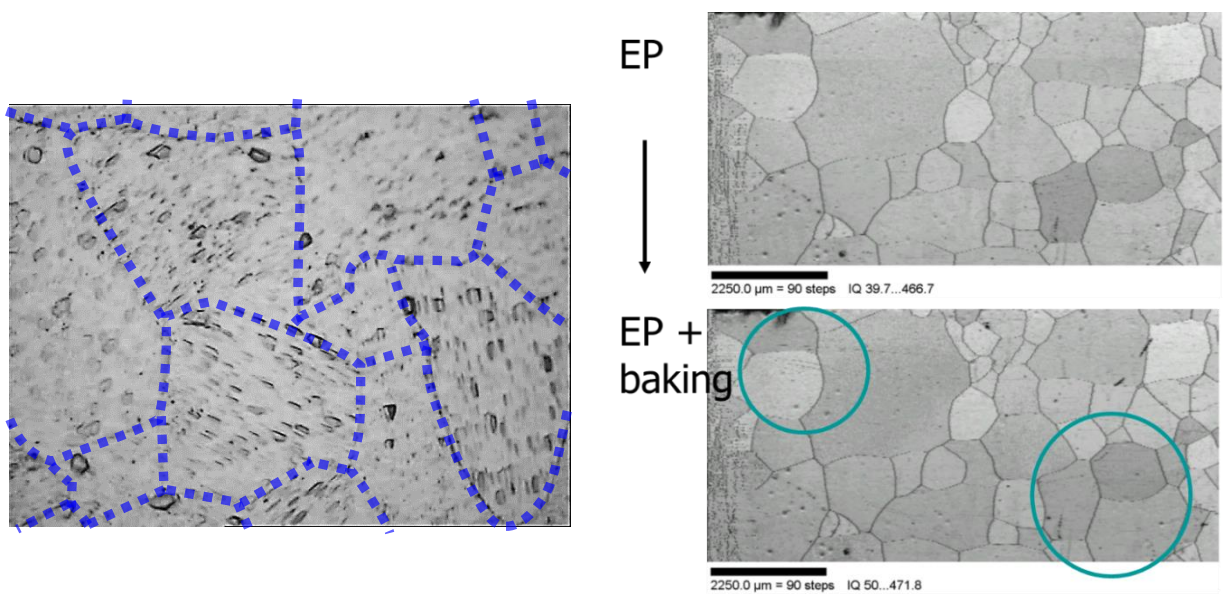

*Figure 4: Left: Etching figures revealing dislocation arrays on as received Nb sheet. Blue dotted line figures grain boundaries. Note how the etching figures obviously depend on crystalline orientation. Right: Crystalline Quality Index as determined by EBSD. The darker areas show grains where the diffraction diagrams are distorted compared to the expected diagram from a "perfect crystal". Note that not all the grains have the same relative quality, and that baking does not necessarily improve the local QI. (only 12-13% of the grains are affected by baking).*

Figure 4 right shows EBSD quality index (QI) of an EP niobium sample before and after baking. EBSD QI is influenced by all types of dislocation (geometrically necessary as well as statistically stored ones), vacancies and impurities. It is an excellent diagnostic of crystalline quality. On this picture one can see that some grains are darker than others (i.e. more "defective"), and that baking affects only some of them. The overall effect is not necessarily and improvement (which can attributed to the fact that baking seems to reduce dislocation density[57] but increases interstitial atoms distribution [58, 59]). Etching figures as shown on Figure 4 left are a way to visualize dislocation cells emerging at the surface of the crystal.

*3.3. Interstitials and vacancies*

Point defects can either be substitutional and interstitials atoms and vacancies. As stated before the pinning potential of point defect is expected to be lower that planar or linear defect, and their effect become sensible only when present in high density. In SRF Nb substitutional atoms are not likely to play a major role: they are minor impurities, and the most abundant of them, Ta, has the same diameter as Nb; it is perfectly miscible in the Nb lattice and does not perturb much the lattice. Interstitials atoms and vacancies on the other hand seem to play a paramount role in Nb.

In [11], Ullmeier *et al* studied the influence of Frenkel pairs (one self-interstitial atom+ one vacancy) created by irradiation below $T_C$ in an nearly perfect Nb wire (reversible magnetization curve). In such conditions they probed defects with relative distance between 20 nm and 4.5 nm, i.e. between 20 and 500 defect per $(\xi)^3 = 10^{-16} cm^3$. As expected, the maximum pinning force depends on the defect density, but not monotonically, which suggest the existence of a fluctuation in the defect density responsible for pinning. Indeed one expects defects to require more energy to be created in e.g. close packed crystalline directions, so irradiation defects (as well as other kind of punctual defects or diffusing species) have no reason to be distributed uniformly at small scale. About the physical origin of the pinning interaction, this paper points out toward to the local modification of the lattice elastic energy. Indeed $\Delta Y/Y$, the relative elastic difference between normal and SC state, is known to decrease with mean free path. Elastic strain of the crystal lattice then interacts with the elastic strain of the vortices lattice. Large quantities of interstitial atoms are known to compress elastically the crystal lattice and influences its SC properties. One well known effect is e.g. the slight enhancement of $T_C$ in doped and/or impure Nb (in strained thin films or see e.g.[60]). In addition, literatures seems to indicate that interstitials play an important role in dislocation behavior upon recovering annealing: although pinning can be observed even in very pure samples, pinning is higher if Niobium with some interstitial content is annealed after a deformation step. Even small amount of interstitial impurities seem to affect the annealed structure of cold worked Niobium [3, 4].

Vacancies also seem to play a paramount role close to the surface. Two authors [61, 62] point out the augmentation of vacancies concentration close to the surface (50 -100 nm) after baking of BCP surfaces. Results seem a little more ambiguous concerning EP surface. Both authors mention the existence of vacancies-H coupling, but their interpretation is radically opposed: Visentin [62] points out the dissociation of those V-H compounds, while Romanenko [61] propose the creation of many new ones via vacancies incorporation from the surface. We think that the first scenario is better supported experimentally as will be discuss below in § 4.1 (Segregations due to crystalline defects). Romanenko also proposed that this release of vacancies favors dislocations movements and is at the origin of the decrease of misorientation angles after baking [63].

*3.4. Conclusion about flux trapping sensitivity.*

Vortex typically interacts with small clusters of a few pinning defects separated by about the coherence length [64]. All data found in literature point out for an effect of fluctuations at the scale of some nm to some 10s of nm. Among the main features liable to be at the origin of flux trapping in SRF cavities, dislocation tangles and/or dislocation cells seem to be the most efficient mechanism since they exist in large number even in well recrystallized Niobium. Obviously large quantity of interstitial can play an additional role, as has been demonstrated with e.g. doping experiments [42, 55]. It is very unlikely that we master yet SRF superconductors, bulk Nb as well as thin films Nb and alternative SC, at this scale.

Recent work done on cavity doping [1] also point out to the influence of interstitials in the flux trapping behavior of Nb cavities. In the next chapter we will deal with their possible effect on the prevention of hydrides precipitation.

## 4. Hydrides:

Hydrides have been identified in the early 90's as source of dissipation upon slow cooldowns [65-71]. Although the bulk concentration was less than 0.01 At.%, precipitation of hydrides was evidenced upon slow cooldown (rate < 1 K/min) if high RRR cavities were allowed to stay between 75-150 K. The quality factor $Q_0$ of slow cooled cavities can lose up to 3 orders of magnitude, and $R_{BCS}$ as well as $R_0$ are affected [67]. The incorporation of H happens principally during etching, but it is a collateral effect resulting from the absence of the oxide which is an effective diffusion barrier to atomic H [69]. In the same way, Niobium with no oxide or depassivated oxide is also very sensitive to H: annealed niobium if kept under vacuum for a long time, Niobium dipped in depassivating solutions (NaOH, HCl, HF…), and even Niobium in contact with boiling water[1] [72]. Hydrogen is known to segregate close to dislocations and form so-called Cottrell clouds [73-76], due to the elastic interaction phenomena described in §3.2 and below.

*4.1. Segregation due to crystalline defects.*

The apparition of hydrides at such low bulk concentration is made possible by the fact that hydrogen, although it is very mobile in Nb, tends to segregate near the surface. This type of segregation is typical of pure metals: indeed hydrogen tend to interact with defects that elastically deform the crystal lattice, the interaction energy increasing from a punctual defect (interstitial atom, vacancy), to a linear defect (dislocation), and to a 2D defect (interfaces like GB, and mostly metal oxide interface). In a sense the defects that tend to pin vortices are also the ones that tend to trap hydrogen atoms, because of the elastic strain they produce on the lattice ([66] and references therein).

In a pure metal most of the defects left are concentrated close to the surface, more exactly to the metal-oxide interface. Measurement show that just underneath the oxide layer the surface concentration of H can reach some 0.1 At.% to some At.% [66, 70]. The oxide-metal interface is a complex area with a lot of strain and dislocations necessary to accommodate the difference in crystalline parameters. Moreover, the surface keeps also traces from the damage layer, mainly produced by rolling and deep drawing of the Nb sheets and even after heavy BCP or EP some grains with a higher density of dislocation still remain [77]. Romanenko showed on cavity cut outs a correlation between hot spot and density of hydrides as well as higher misorientation angles [63]. Hydrides have also been evidenced by Raman with characteristic vibrational bands from NbH and $NbH_2$ have been observed on cold as well hot spots [78], but their density seems to be higher on hot spots.

Note that some of the hydrogen can also be trapped by vacancies. Two situations may occur: in a single vacancy up to 6 H can be trapped, and they are evidences that more H tends to gather in the vicinity, providing nucleation site for hydrides [79]. In presence of interstitials atoms like O, N, C, those atoms will occupy preferentially the vacancy: each oxygen, nitrogen or carbon impurity traps only a single hydrogen atom under formation of oxygen-hydrogen, nitrogen-hydrogen or carbon-hydrogen pairs, where the trapped hydrogen occupies interstitial trap sites in the neighborhood of the trapping impurity [80-83]. Specific interaction between interstitial oxygen and hydrogen has been thoroughly studied in niobium [76, 81, 82, 84-88] as it constitute a textbook example of two-well tunneling system.

There are direct evidences that hydride precipitation is inhibited due to trapping at O,N or C interstitials [80, 89]. Moderate heating like baking is known to break the interstitial-hydrogen bond [75, 83, 87]. It has also been proven that the near surface is enriched into interstitials, mainly O [53, 66], and that baking tends to increase the O content in the near surface [59, 90]. We can thus infer a scenario explaining in the same time the baking effect and doping experiments: if the interstitial content near the surface (~RF penetration depth), is increased, the interaction of H with those interstitials will prevent accumulation of hydrogen and reduce the apparition of hydrides. Indeed recent TEM experiment show the decreased hydride formation after 120 °C baking [91]. Hydrides being poor superconductors, they can promote early penetration of vortex, thus by reducing their presence, one can reach higher fields without early nucleation of vortex, which is basically what is observed after baking. The initial absence of vortex is sustained by µ-SR experiment showing that the magnetic volume is ~ 0% on cold spots until fields close to the quench field [24]. Another indication of the good crystalline quality of "cold spots" vs hot spot is given by tunneling PCT experiment: cold spot exhibit a nearly ideal BCS behavior, while hot spots show enlarged Density of State, localized energy level inside the gap and other features relevant from inelastic pair breaking [92].

*4.2. Heat treatments: necessary but not fully efficient.*

Heat treatment (UHV, typically 800°C, 2h, but also 600°C 5-10 h) allow to get rid of most of the heavy Q-disease. From the thermodynamic point of view, H is meant to degas, but the kinetic is controlled by diffusion [93], and some hydrogen keeps trapped inside the Nb matrix. The effect of annealing is multiple: as mentioned before, it will dissociate vacancies-H binds, it will render the distribution of H more or less uniform in the material (as long as it keeps warm) and in the same time, will provide some restauration/recrystallization of the material. Upon cooling, the remaining hydrogen tends to segregate again toward the surface [66, 94]. Nevertheless, annealed cavities are much less sensitive to hydrogen uptake during subsequent chemical polishing. This is probably related to the fact that the annealing allows the lattice to recrystallize and/or recover, and thus the quantity of crystalline defects, in particular dislocations are tremendously reduced, leaving less preferential sites for segregation [95, 96]. One argument that support further the recrystallization hypothesis, is that the couples "times + temperature" observed to remove Q-disease are typically those known for recrystallization: the relationship between time ($t_R$) and temperature ($T_R$) to achieve a given recrystallization state R is in the form of [97]:

$$\log t_R = A + B\left(\frac{1}{T_R}\right) \quad (1)$$

where time is in hours and temperature in Kelvins. A and B depend on the purity and deformation state prior annealing.

---

[1] Note that typically degreasing is conducted in slightly basic detergent solution at around 50-60°. Survey of this step was never closely documented, although it could be at the origin of some of the H pollution, basic detergent solution being depassivating and water at high temperature known to be harmful.

For a typical Nb batch, it was determined that the law was [53], but it probably needs to be estimated precisely for each batch of material:

$$\log t_R = 25861 \cdot \left(\frac{1}{T_R}\right) - 22.28 \quad (2)$$

Table 1 : typical annealing time needed for recrystallization of Nb at various temperature

| T | T K | Tcalc (h) |
|---|---|---|
| 500 | 773 | 11,18 |
| 600 | 873 | 7,34 |
| 675 | 948 | 5,00 |
| 700 | 973 | 4,30 |
| 750 | 1023 | 3,00 |
| 800 | 1073 | 1,82 |
| 850 | 1123 | 0,75 |

Nevertheless, even if heat treatment allow to get rid of heavy Q-diseases, experimental facts described in the previous sections indicate that it is not possible to eliminate all the crystalline defects, and some hydrogen remains. They are some indication that different kind of hydrides are in concern depending on the H content is low or high and on the crystalline state of the surface [95].

**5. Flux trapping vs flux early entry.**

Hotspots are attributed to the presence of vortices [98, 99], but those vortices can have two possible origins: initially trapped flux line, or early flux penetration when increasing RF field. As stated before any defects with depleted superconductivity is liable to promote early vortex entry as well as pinning behavior. Pinning force also depends on field, and it is possible that the predominant kind of defects is not the same in the case of remnant flux trapping compared to early field penetration on RF. They are several experimental evidences that support each phenomena and they probably coexist in many cases.

The patchy nature of hot spot has been established quite early [38], but it until the thorough work from A. Romanenko on cavity cuts that a better understanding appeared [57, 100]. A. Romanenko and collaborators worked on samples cut out of cavities (small grain or large grain, BCP or EP) and compared cold and hot spots from the structural, morphological and chemical point of view [4, 5, 17, 18]. Recently these results have been completed with magnetic characterization that highlight some aspects of vortices pinning [24, 101].

It ensued that the main difference between cold and hot spot was observed by electron diffraction technique that probes the first ~ 100 nm of the sample (EBSD, also use for orientation imaging). Indexation of the diffraction pattern give the exact orientation of each grain, but can also give access to an evaluation of the local misorientation and indirectly to dislocation density. Romanenko observed that hot spots tend to exhibit higher misorientation angles and thus higher dislocation density[2]. He also found that Nb cut off samples, after a metallographic mechanical polishing, exhibit characteristic features of hydrides precipitates (snowflake like features highly visible on cold SEM pictures), and that the density of hydrides is higher on hot spots than it is on cold spots [102, 103]. Similar hydrides visualization technique has also been used in [104]. It is fairly obvious that high dislocation density and the presence of high concentration of hydrides are related.

As stated earlier, hot spots can either be attributed to initially trapped flux line at cooldown of cavities or to early flux penetration with increasing RF field. In Table 2: trapped vortices vs early penetration: experimental arguments, extracted from [19, 24, 25, 53, 101, 105, 106] we list the experimental arguments in favor of one or the other hypothesis. It is highly probable that both phenomena co-exist.

---
[2] Note that with the misorientation technique used here, one is sensitive only to the geometrically necessary dislocation resulting from lattice deformation. It does not give information on the overall crystalline disorder (equilibrium density of dislocation, vacancies, impurities…).

In particular μSpin rotation experiments are very useful as they allow to probe altogether the local magnetic field and what proportion of the volume is affected. Experiments on cavity cut-outs in [24] (and ref therein) show that DC ⊥ magnetic field starts to enter Nb around ~100 mT at 2.5 K (where pure DC BCS behavior should give 170 mT). What is more surprising is that enhanced surface dissipation appeared at the exact same field in RF // field on the cavity before it was cut. So we obviously deal with the some common phenomenon (note that the paper also proposes surface roughness to be at the origin at these shared properties, as it produces perpendicular components of the field even in parallel field).

*Table 2: trapped vortices vs early penetration: experimental arguments, extracted from [19, 24, 25, 53, 101, 105, 106]*

| Trapped vortices | Early penetration |
|---|---|
| • μSR : without mask, threshold @ 40 mT but inner volume essentially non magnetic:could be related to weak pinning on edges, surface. | • μSR: before 100 mT « volume magnetic fraction » is ~0% for masked samples (ie without edge effect) : no flux line originally present<br>• at H> 100 mT, lines jump directly to the center of the samples |
| • Irreversibility observed in DC magnetization curve<br>• Flux jumps on some magnetization curves (NB. Flux jumps are more often observed on damaged or polluted surfaces).<br>• ~ 0 magnetic moment on field cooled DC experiments | • Point contact Tunneling shows degraded SC properties in hot spot (inelastic quasiparticle scattering: broadened DOS, a finite zero bias conductance and/or zero bias peak, lower gap…) : could promote early penetration |
| • Depining frequency is very low for bulk Nb ($10^4$ Hz) : trapped flux is expected to be mobile in RF field and produce dissipation | • Morphologic defects are often found at the quench site (step, pit): in this case the local field is higher than the applied field. It can also be considered as having a defect with locally lower $B_{C1}$ |

It is interesting to note that Nb thin films show common properties with bulk niobium hotspots, in particular the existence of sub gap states as measured with point contact tunneling [101], since we know that thin films usually exhibit high crystalline defect density: strained lattice with higher $T_C$, very small grains, high density of displaced and foreign atoms.... Similar problems are also expected to be encountered for alternative SRF superconductors like $Nb_3Sn$, $MgB_2$, multilayers…

**6. Conclusion:**

It is interesting to note that crystalline state monitor two of the most harmful phenomena limiting SRF performance: flux trapping and early flux penetration due to hydrides.

Even if one cannot get a perfect surface in realistic conditions there is still room for optimization of cavity preparation process. Dealing with bulk Nb, during the fabrication, several aspects are not followed closely enough to get reproducible quality of niobium within the SRF depth of interest [53]:

- Recrystallization: during Nb sheet fabrication, several steps of heavily deformation + recrystallization stages occur. Firstly, depending on the desired thickness, the deformation ratio is not exactly the same for all the sheets. Secondly, depending on their position in oven, not all sheets undergo the recrystallization treatment at the same exact temperature. It follows that the dislocations distribution in the material can vary a lot within a same batch, and even more from batch to batch or vendor to vendor.
- Surface damage due to rolling, and specifically skin pass is known to affect a deeper thickness than the 150 μm usually chemically withdrawn on the cavities surface. As pickling of the sheets is generally applied at the vendor and removes an unknown amount of damage layer, the actual damage thickness to be removed is not known with precision. This aspect also contribute to batch to batch variation.
- Forming and welding of cavities also contribute to additional strain. In particular, welding is somehow equivalent to a heat treatment and produce some recovering, but if not conducted correctly, it can also give rise to local thermal stresses in the heat affected zone because of the huge thermal gradient proper to EB welding. This remaining stress can be at the origin of local stress corrosion and the apparition of pits during subsequent etching.
- EP/BCP and/or CBP of cavities has a benefic effect as it allows removing most of the damage layer, but here again, some grains more defective than the adjoining ones may persists if not enough material is removed.
- The baking and heat treatments usually applied after heavy EP/BCP of cavities have a benefic effect as they allow recovering a large portion of crystal defects, but since there is not a universal recipe among different labs, it is also a source of non-reproducibility. Moreover, if these treatment seem to be beneficial in term of crystal structure, they tend to increase the interstitial content, and might replace a strong localized pinning with multiple weak pinning centers (magnetic to volume pinning). Determining the proper balance between various treatments in function of desired performance will be for sure the next challenge in SRF.

The crystallographic quality required for SRF application is beyond the reach of traditional metallurgical characterization. Ultimately it would be beneficial to systematically assess the surface state of representative coupons (which underwent similar process as the whole cavity fabrication) with advanced techniques like EBSD QI, PCT, µSR, etc… to have a better understanding and predictability of the batch to batch, cavity to cavity variations. Even if it is expensive, cutting actual cavities has proven an infinite source of information and some provisions should be made to go on in this direction.